\documentstyle[prl,aps,preprint,tighten,floats,epsf]{revtex}

\newbox\rotbox

\begin{document}
\draft
\setcounter{page}{0}
\def\footnoterule{\kern-3pt \hrule width\hsize \kern3pt}

\preprint{\vbox{
                                        \null\hfill\rm MIT-CTP-2678\\
                                        \null\hfill\rm U.of MD PP\# 98041}}
%
\title{Nucleon sigma term and in-medium quark condensate in the modified 
quark-meson coupling model\thanks
{This work is supported in part by funds provided by the U.S.
Department of Energy (D.O.E.) under cooperative 
research agreement \#DF-FC02-94ER40818.}}
\author{Xuemin Jin$^1$ and 
Manuel Malheiro$^2$\thanks{%
Permanent address: Instituto de F\'{\i}sica, Universidade Federal
Fluminense, 24210-340, Niter\'oi, R. J., Brasil.}}
\address{$^1$Center for Theoretical Physics,
Laboratory for Nuclear Science
 and Department of Physics\\
Massachusetts Institute of Technology
Cambridge, Massachusetts 02139, USA\\
$^2$Department of Physics,
 University of Maryland,
College Park, MD 20742, USA} 
%
%
\maketitle

\thispagestyle{empty}
%
\begin{abstract}

We evaluate the nucleon sigma term and in-medium quark condensate in the
modified quark-meson coupling model which features a density-dependent 
bag constant. We obtain a nucleon sigma term consistent with its 
empirical value, which requires a significant reduction of the bag 
constant in the nuclear medium similar to those found in the previous 
works. The resulting 
in-medium quark condensate at low densities agrees well with the model 
independent linear order result. At higher densities, the magnitude of 
the in-medium quark condensate tends to increase, indicating no tendency
toward chiral symmetry restoration.  

\end{abstract}
\vspace*{\fill}
\begin{center}
Published: {\it Mod. Phys. Lett. A,} Vol. 14, No. 4 (1999) pp. 289-297
\end{center}

\narrowtext
%

The physics of nuclear matter and finite nuclei is governed by the 
underlying theory of strong interactions of quarks and gluons,  
quantum chromodynamics (QCD).  In reality, however,  QCD is intractable 
at the nuclear physics energy scales due to the nonperturbative features 
of the theory and a realistic account of nuclear phenomena based
entirely on it is not yet possible. At this stage, the best one 
could do is to build models that incorporate the symmetries of QCD
and/or quark-gluon degrees of freedom, and hence to motivate connections 
between this theory and the observed nuclear phenomena and established 
phenomenology. Such models are necessarily quite crude.

The quark-meson coupling model (QMC) \cite{guichon88} treats nucleons
in nuclear medium as non-overlapping MIT bags interacting through 
the self-consistent exchange of mesons in the mean-field approximation. 
It provides a simple and attractive framework to incorporate the quark 
structure of the nucleon in the study of nuclear 
phenomena \cite{guichon88,fleck90,%
saito94,saito95,saito-gen,blunden96}. 
Recently, the QMC has been modified by introducing a density-dependent 
bag constant \cite{jin96,jin96a}. It was demonstrated that a significant 
reduction of the bag constant in the nuclear matter relative to its free 
space value can lead to large and canceling isoscalar Lorentz scalar and 
vector potentials and hence strong spin-orbit force for the nucleon in 
nuclear matter which are comparable to those suggested by relativistic 
nuclear phenomenology \cite{wallace87,serot86} and finite-density QCD sum 
rules \cite{cohen95}. Such a large reduction of the bag constant can
also account for the EMC effect within the framework of dynamical 
scaling \cite{jin97}. (For further development and other applications,
see Refs.~\cite{muller97,panda97}).

In this paper, we evaluate the nucleon $\sigma$ term and in-medium quark
condensate in the modified quark-meson coupling model (MQMC) developed 
in Refs.~\cite{jin96,jin96a}. The in-medium quark condensate  
plays an important role in studying hadron properties in nuclear 
medium~\cite{cohen95} and has connections to many nuclear 
phenomena \cite{adami93}. Its value at low densities is largely 
determined by the nucleon $\sigma$ term which also has attendant 
consequences in nuclear astrophysics \cite{kaplan86}.
In Refs. \cite{saito94,saito95} both the nucleon $\sigma$ term and 
the in-medium quark condensate have been investigated in the QMC. There,
it was found that the result for the nucleon $\sigma$ term is much smaller 
than its empirical value extracted from dispersion analysis of isospin
even pion-nucleon scattering \cite{gasser91}, and the prediction for 
the in-medium quark condensate based on this small $\sigma$ term is quite 
different from the model independent linear order result 
\cite{drukarev90,cohen92,lutz92}. It is thus of interest to examine 
how the large reduction of the bag constant in the nuclear matter suggested 
in the MQMC will affect the nucleon $\sigma$ term and in-medium quark 
condensate.

We find that when the bag constant is significantly reduced in the nuclear 
matter, e.g. $B/B_0 \sim 35-40\%$ at the nuclear matter saturation density, 
the empirical value for the nucleon $\sigma$ term can be recovered
and the resulting in-medium quark condensate at low densities is 
in good agreement with the model independent linear order result. At high
densities, the magnitude of the in-medium quark condensate tends to increase, 
indicating no tendency of chiral symmetry restoration; this behavior has also 
been seen in other models \cite{li94,brockmann96,delfino95,malheiro97}. 
Such a large reduction of the bag constant, 
as shown in previous works \cite{jin96,jin96a,jin97}, is consistent with
that required to recover large and canceling Lorentz scalar and vector 
potentials for the nucleon in the nuclear matter and to account for the EMC
effect within the dynamical rescaling framework. 

The details of the MQMC have been given in \cite{jin96,jin96a}.
The two models for the in-medium bag constant are:
 scaling model and direct coupling model. The scaling 
model relates the in-medium bag constant to the in-medium nucleon mass : 
\begin{equation}
{B\over B_0} = \left[ M_N^*\over M_N \right]^\kappa\ ,
\label{an-br}
\end{equation}
where $\kappa$ is a real positive parameter and $\kappa=0$ corresponds 
to the usual QMC model.
The effect of this modification is summarized
into a factor $C(\overline{\sigma})$  that appears
in the self-consistency condition for the $\sigma$ field
\cite{jin96,jin96a},
\begin{equation}
C(\overline{\sigma}) = {E_{\rm bag}\over M_N^*}
\Biggl[\left(1-{\Omega_q\over E_{\rm bag}\, R}\right)\,
S(\overline{\sigma}) +
{m_q^*\over E_{\rm bag}}\Biggr]
\Biggl[1-\kappa\, {E_{\rm bag}\over M_N^{* 2}}
{4\over 3}\,\pi\,R^3\, B\,
\Biggr]^{-1}\ ,
\label{tc-k}
\end{equation}
where $m^*_q = m_q - g^q_\sigma\overline{\sigma}$, and the explicit 
expressions for $M^*_N, E_{\rm bag}, \Omega_q$, and $S(\overline{\sigma})$
can be found in Refs.~\cite{jin96,jin96a}.
The expression for $C(\overline{\sigma})$ in the usual QMC model is obtained
from Eq.~(\ref{tc-k}) with $\kappa=0$.
The direct coupling model features a direct coupling between the bag 
constant and the scalar mean field
\begin{equation}
{B\over B_0} = \left[ 1
- g_\sigma^B\, {4\over \delta} {\overline{\sigma}\over M_N} \right]^\delta\ ,
\label{an-dir}
\end{equation}
where $g_\sigma^B$ and $\delta$ are real positive parameters and 
the introduction of $M_N$ is based on the consideration of dimension.
(The case $\delta = 1$ was also considered by 
Blunden and Miller \cite{blunden96}.)
Note that $g_\sigma^B$ differs from the quark-meson 
coupling $g_\sigma^q$ (or $g_\sigma\equiv 3g_\sigma^q$).
When $g_\sigma^B = 0$, the usual QMC model is recovered.
The factor $C(\overline{\sigma})$, in this case, is given by
\begin{equation}
C(\overline{\sigma}) =
{E_{\rm bag}\over M_N^*}
\Biggl[\left(1-{\Omega_q\over E_{\rm bag}\, R}\right)\,
S(\overline{\sigma}) +
{m_q^*\over E_{\rm bag}}\Biggr] + 
\left({g_\sigma^B\over g_\sigma}\right)\,
{E_{\rm bag}\over M_N^*}\,
{16\over 3}\,\pi\, R^3\,{B\over M_N}\,
\Biggl[1-{4\over \delta}{g_\sigma^B \overline{\sigma}\over M_N}
\Biggr]^{-1}\ .
\label{tc-d}  
\end{equation}

The nucleon $\sigma$ term can be expressed as
\begin{equation}
\sigma_N = m_q {d M_N\over d m_q} 
= 2 m_q\langle N|\overline{q}q|N\rangle\ .
\label{sterm-def}
\end{equation}
Here we neglect isospin breaking and use 
$m_u=m_d \equiv m_q = {1\over 2}(m_u+m_d)$. 
Therefore, there are two ways to evaluate $\sigma_N$ in
QCD. One is to take the derivative of $M_N$ with respect to $m_q$.
The other is to calculate the nucleon's scalar charge 
$\langle N|\overline{q}q|N\rangle$, which can be carried out by 
adding a term $S (\overline{u} u +
\overline{d}d)$ (with $S$ a constant) to the QCD Hamiltonian and
then extracting the response of the nucleon mass to the external
field $S$, $dM_N(S)/dS|_{S\rightarrow 0} = \langle N|
\overline{u}u+\overline{d}d|N\rangle$.
The QMC and MQMC provide descriptions of how the nucleon
mass responds to a constant scalar field. Treating 
$g_\sigma^q\overline{\sigma}$ as a constant external field,
one can show
\begin{equation}
\sigma_N =  m_q {dM_N^*(S=-g_\sigma^q \overline{\sigma})
\over d S}\mid_{S\rightarrow 0} 
= 3 m_q C(\overline{\sigma}=0)\ .
\label{sterm-cal}
\end{equation}
where $C(\overline{\sigma}=0)$ is related to the response of the
nucleon mass to the scalar field at $\sigma = 0$. The
explicit expressions for $C(\overline{\sigma})$ are in the
Eqs.~(\ref{tc-k}) and (\ref{tc-d}). 

\begin{figure}[b]
\begin{minipage}[h]{6.0in}
\vspace*{-3cm}
\epsfxsize=7.0truecm
\centerline{\epsffile{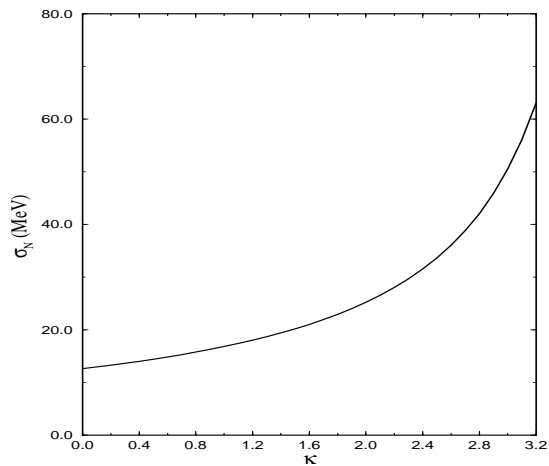}}
\vspace*{1cm}
\caption{Nucleon $\sigma$ term as a function of 
$\kappa$ resulting from the scaling model. The case $\kappa=0$
corresponds to the usual QMC.}
\label{fig-1}
\end{minipage}
\end{figure}

\begin{figure}[t]
\begin{minipage}[h]{6.0in}
\vspace*{-3cm}
\epsfxsize=7.0truecm
\centerline{\epsffile{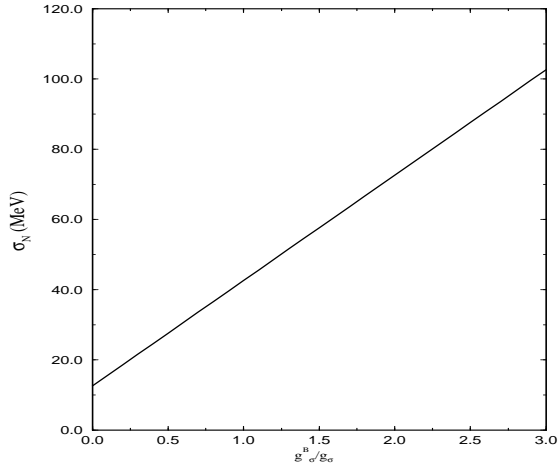}}
\vspace*{1cm}
\caption{Nucleon $\sigma$ term as a function of 
$g^B_\sigma/g_\sigma$ resulting from the direct coupling model.
The case $g^B_\sigma/g_\sigma = 0$ corresponds to the usual QMC.}
\label{fig-2}
\end{minipage}
\end{figure}

We follow Refs.~\cite{saito94,saito95} and take a value of $10$ MeV for 
$m_q$. We expect this value to be reasonable at the scale where the MIT
bag model is useful. Note that $C(0)$ also depends on $m_q$. Here we use $m_q=0$ in 
evaluating $C(0)$. It has been found in previous studies that inclusion of a small 
finite $m_q$ in evaluating $C(0)$ only leads to a negligible refinement 
to the $C(0)$ value. Figure~\ref{fig-1} shows
the resulting $\sigma_N$ from the scaling model as a function of $\kappa$
for $R_0 = 0.6$ fm. When $\kappa=0$, corresponding to the usual QMC,
$\sigma_N \simeq 12$ MeV which is almost a factor of four smaller than
the empirical value of $45$ MeV \cite{gasser91}. As $\kappa$ increases, 
$\sigma_N$ increases slowly at small $\kappa$ values and grows rapidly 
at large $\kappa$ values.
For $\kappa\simeq 2.88$, we find $\sigma_N\simeq 45$ MeV. We also find that 
the result is largely independent
of $R_0$ in the range 0.6 fm $\leq R_0\leq 1.0$ fm.
The result from the direct coupling model is illustrated in Fig.~\ref{fig-2},
where $\sigma_N$ is plotted as a function of $g^B_\sigma/g_\sigma$ with
$R_0 = 0.6$ fm.  We see that 
the dependence of $\sigma_N$ on $g^B_\sigma/g_\sigma$ is linear. 
The case $g^B_\sigma/g_\sigma = 0$
corresponds to the usual QMC. When $g^B_\sigma/g_\sigma \simeq 1.1$, the 
empirical value of $\sigma_N$ can be reproduced. Our results for 
$\kappa$ in the scaling model and $g^B_\sigma/g_\sigma$
in the direct coupling model in order to obtain the empirical value of
$\sigma_N$ term depend slightly on the choise we make for $m_q$ in the 
vicinity of $m_q\sim 10$MeV.

Therefore, it is necessary to have $\kappa>0$
and $g^B_\sigma/g_\sigma>0$ or a {\it reduction} of 
the bag constant in the nuclear medium in order to recover 
the empirical value of $\sigma_N$ from the small value predicted 
by the usual QMC. If one had assumed the opposite behavior,
the resulting $\sigma_N$ would be even smaller. This  
solidifies the MQMC model.  It is the coupling of 
the bag constant to the scalar field featured in the MQMC that 
describes how the bag constant responds to the external field and 
provides a new source of contribution 
to the nucleon's scalar charge and hence the nucleon $\sigma$ term. 

For $\kappa \sim 2.9$ (in the scaling model) and 
$g^B_\sigma/g_\sigma \sim 1.1$ (in the direct coupling model), 
$B/B_0 \simeq 35-40\%$ at the saturation density \cite{jin96a}. 
It is rewarding that such a significant reduction of the bag 
constant in the nuclear matter relative to its free-space value
{\it coincides} with that required to recover the relativistic
nuclear phenomenology and account for the EMC effect \cite{jin96,jin96a,jin97}.
Conversely, the empirical value $\sigma_N = 45$ MeV provides 
an extra constraint to the MQMC. In fact,  there will be no free 
parameter in the scaling model and only $\delta$ is left as a 
free parameter in the direct coupling model when $\sigma_N$
is treated as an input. Then the resulting scalar and vector 
potentials for the nucleon in the nuclear matter will be 
comparable to those suggested by the relativistic nuclear 
phenomenology and finite-density QCD sum rules, and the predictions 
for the rescaling parameter will be consistent with that required
to explain the EMC effect within the dynamical rescaling 
approach.

We now turn to the in-medium quark condensate.
Following Ref. \cite{cohen92}, we can write the in-medium
quark condensate as
\begin{equation}
 \langle\overline{q}q\rangle_{\rho_N}
= \langle\overline{q}q\rangle_0 
+{1\over 2} {d{\cal E}\over dm_q}
= \langle\overline{q}q\rangle_0 +{1\over 2}\Biggl(
{\partial {\cal E}\over \partial M_N^*} {d M_N^*\over d m_q}
+ \chi_\sigma {\sigma_N\over m_q} 
{\partial {\cal E}\over \partial m_\sigma}
+\chi_\omega {\sigma_N\over m_q} 
{\partial {\cal E}\over \partial m_\omega}\Biggr)
\label{e-der}
\end{equation}
where ${\cal E}$ is the energy density of the nuclear medium, 
$\chi_\sigma \sigma_N/ m_q\equiv d m_\sigma/ d m_q$, 
$\chi_\omega \sigma_N/ m_q\equiv d m_\omega/ d m_q$, 
%
and  $\langle\overline{q}q\rangle_{\rho_N}$
and $\langle\overline{q}q\rangle_0$ denote the quark condensates
in the nuclear medium and vacuum, respectively. 
Here we have followed Ref.~\cite{cohen92} and neglected
the dependence of various couplings on $m_q$.
Using the Gell-Mann--Oakes--Renner relation,
$ 2 m_q \langle\overline{q}q\rangle_0 = -m_\pi^2 f_\pi^2$,
one finds \cite{saito94}
\begin{equation}
R_\rho \equiv {\langle\overline{q}q\rangle_{\rho_N}\over 
\langle\overline{q}q\rangle_0 }
= 1 - {\sigma_N \rho_N\over m_\pi^2 f_\pi^2}
\Biggl[ {m_\sigma^2 \overline{\sigma} \over
g_\sigma C(0) \rho_N} + \chi_\sigma {m_\sigma 
\overline{\sigma}^2\over \rho_N}
-\chi_\omega {g_\omega^2\over  m_\omega^3}\rho_N
\Biggr]\ ,
\label{cond-eva}
\end{equation}
where $m_\pi$ is the pion mass (138MeV) and $f_\pi$ the pion decay constant
(93 MeV). Here we have used $dM_N^*/dm_q = 3 m_q C(\overline{\sigma})$
\cite{saito94,saito95},
which can be obtained by following the same discussion leading to 
Eq.~(\ref{sterm-cal}). 

\begin{figure}[t]
\begin{minipage}[h]{6.0in}
\vspace*{-3cm}
\epsfxsize=7.0truecm
\centerline{\epsffile{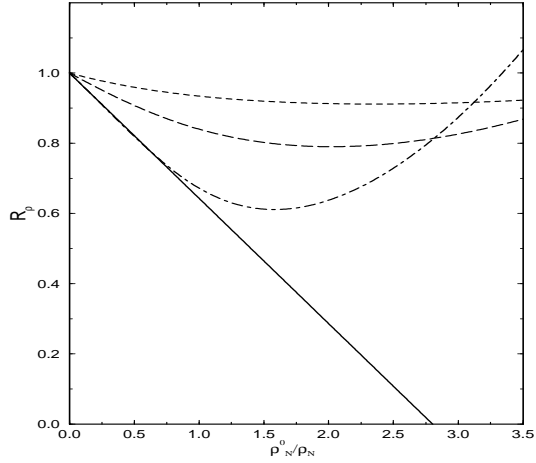}}
\vspace*{1cm}
\caption{Ratio $R_\rho = \langle\overline{q}q\rangle_{\rho_N}/
\langle\overline{q}q\rangle_0$ as a function of the medium density
from the scaling model with $R_0 =0.6$ fm. 
The $\sigma_N$ value is predicted from
Eq.~(\protect\ref{sterm-cal}). The solid curve represents the 
linear order result with $\sigma_N=45$ MeV \protect\cite{cohen92}, 
 and the other three curves correspond to $\kappa =0$ (dashed) 
(usual QMC, $\sigma_N\simeq 12.6$ MeV), 2.0 (long-dashed)
 ($\sigma_N\simeq 25.2$ MeV),
and 2.88 (dot-dashed) ($\sigma_N\simeq 45.0$ MeV), respectively.} 
\label{fig-3}
\end{minipage}
\end{figure}

To be self-consistent, we use the predictions for $\sigma_N$ from 
Eq.~(\ref{sterm-cal}). (In Refs.~\cite{saito94,saito95}, the empirical value of
$\sigma_N$ was used in evaluating the in-medium quark condensate.)
The two coupling constants,
$(g_\sigma,g_\omega)$ for the scaling model and $(g^B_\sigma,
g_\omega)$ for the direct coupling model, are chosen to fit
the nuclear matter binding energy  ($-16$ MeV)
at the saturation density ($\rho_N^0 = 0.17$ fm$^{-3}$). Here we also 
adopt the ansatz $\chi_\sigma\simeq m_\sigma/M_N$ and
$\chi_\omega\simeq m_\omega/M_N$ suggested in Ref.~\cite{cohen92}
with $m_\sigma = 550$ MeV and $m_\omega = 783$ MeV.
The predictions from the scaling model are presented in Fig.~\ref{fig-3}, 
where $R_\rho$ is plotted as a function of the nuclear medium density 
for different $\kappa$ values, with $R_0 = 0.6$ fm. For comparison, 
the model independent linear order result with $\sigma_N = 45$ MeV, 
obtained by using ${\cal E} = \rho_N M_N$ 
($R_\rho = 1 - \sigma_N\rho_N/m_\pi^2 f_\pi^2$) \cite{cohen92}, 
is also shown. For $\kappa$ values which we are interested here,
 $R_\rho$ decreases with density 
at low densities and 
then starts to increase at higher densities. We see that
the dependence of $R_\rho$ on $\rho_N$ is almost linear at
low densities with the slope essentially determined by the
$\sigma_N$ value. When $\kappa\simeq 2.88$ ($\sigma_N\simeq 45$ MeV), 
the result in the region $\rho_N < 1.2\, \rho_N^0$ agrees well 
with the linear order result and starts to grow quickly 
at $\rho_N \sim 1.8\, \rho_N^0$. For smaller (larger) $\kappa$ values, 
the result at low densities becomes larger (smaller) than the linear order 
result, which can be attributed to smaller (larger) $\sigma_N$ values.
The rate of the increase of $R_\rho$ with density at higher densities
increases with increasing $\kappa$ value.

\begin{figure}[t]
\begin{minipage}[h]{6.0in}
\vspace*{-3cm}
\epsfxsize=7.0truecm
\centerline{\epsffile{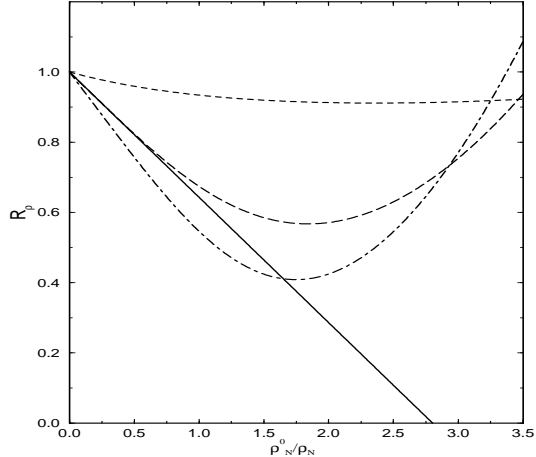}}
\vspace*{1cm}
\caption{Ratio $R_\rho = \langle\overline{q}q\rangle_{\rho_N}/
\langle\overline{q}q\rangle_0$ as a function of the medium density
from the direct coupling model with $\delta=4$ and $R_0=0.6$ fm. 
The $\sigma_N$ value is predicted from
Eq.~(\protect\ref{sterm-cal}). The solid curve represents the 
linear order result with $\sigma_N=45$ MeV \protect\cite{cohen92}, 
and the other three
curves correspond to $g^q_\sigma=5.309$ (dashed) 
(usual QMC, $\sigma_N\simeq 12.6$ MeV), 
1.93 (long-dashed) ($\sigma_N\simeq 45.0$ MeV), and 
1.5 (dot-dashed) ($\sigma_N\simeq 61.2$ MeV), respectively.}
\label{fig-4}
\end{minipage}
\end{figure}

The resulting $R_\rho$ from the direct coupling model is plotted 
in Fig.~\ref{fig-4}. The results are for $\delta =4$ and different
values of $g^q_\sigma$.  For $g^q_\sigma \simeq 1.93$
($\sigma_N \simeq 45$ MeV), the prediction for $R_\rho$ is in good 
agreement with the linear order result below $1.2\, \rho_N^0$; at
$\rho_N\sim 1.9\,\rho^0_N $, $R_\rho$ starts to increase with density. As
$g^q_\sigma$ gets larger (smaller), the $\sigma_N$ becomes smaller
(larger) and hence $R_\rho$ becomes larger (smaller). At higher
densities, $R_\rho$  increases with increasing density
with a large (small) rate for small (large) $g^q_\sigma$ values.
We also tested the sensitivity to the $\delta$ value. For a given 
$g^q_\sigma$, the result is not sensitive to $\delta$ value in the 
regime $\rho_N < 2.5 \rho^0_N$, and saturates at $\delta\sim 12$. 

We observe that the ratio $R_\rho$ at low densities
($\rho_N\leq \rho_N^0$) is essentially determined by the
$\sigma_N$ value and the linear order result is robust. 
This is consistent with that found in the usual QMC \cite{saito94,saito95}
and in hadronic models \cite{cohen92,li94,brockmann96,delfino95,malheiro97}.
At higher densities, the nonlinear higher-order contributions
become increasingly important. In particular, the last term
in large parentheses in Eq.~(\ref{cond-eva}) becomes dominant, leading to a
hindrance of chiral symmetry restoration. This behavior
is also seen in hadronic models 
\cite{li94,brockmann96,delfino95,malheiro97}. 
An exception \cite{delfino95,malheiro97} is the
ZM model \cite{zimanyi90}, which features density-dependent
meson-nucleon coupling constants. However, since the parameters in 
the MQMC and in hadronic models are chosen to fit only the nuclear
matter properties at the saturation density, the reliability of 
their predictions for the in-medium quark condensate at high densities
is unknown. Moreover, the high-density behavior of $R_\rho$ is sensitive to 
$\chi_\sigma$ and $\chi_\omega$, which are not well determined
\cite{brockmann96}. The possible dependence of the couplings, 
$g_\sigma^q$, $g_\sigma^B$, and $g_\omega$, on $m_q$ is also
neglected, which would give extra contributions to $R_\rho$.

In summary, we have evaluated the nucleon $\sigma$ term and 
in-medium chiral quark condensate in the modified quark-meson
coupling model. The coupling of the bag constant to the scalar
mean field featured in the MQMC gives rise to additional contribution
to the nucleon $\sigma$ term compared to the usual QMC. This
contribution can lead to the recovery of the empirical value
of the nucleon $\sigma$ term when the reduction of the bag constant
in the nuclear matter relative to its free space value is
significant, i.e., $B/B_0 \simeq 35-40\%$ at $\rho_N=\rho^0_N$.
Such a large reduction of the bag constant in the nuclear matter is 
consistent with that required to recover large and canceling Lorentz 
scalar and vector potentials for the nucleon in the nuclear matter 
and to account for the EMC effect within the dynamical rescaling framework.
The resulting in-medium quark condensate 
at low densities agrees well with the model independent linear order 
result; at higher densities, the magnitude of the in-medium quark 
condensate tends to increase, indicating no tendency toward chiral 
symmetry restoration.

\vspace*{0.5cm}
One of us (MM) would like to thank the TQHN group at the University
of Maryland for their hospitality during his extended visit, and 
the Brazilian agency CAPES Grant No. BEX1278/95-2 for the financial 
support which made this visit possible. 

\end{document}